\documentclass[graybox, envcountchap]{svmult}

\usepackage{graphicx}        
\graphicspath{
	{book/}
	{book/chapter-introduction/}
	}

\usepackage{multicol}        
\usepackage[bottom]{footmisc}
\usepackage{enumitem}

\usepackage{newtxtext}       %

\usepackage{url}
\usepackage{pifont}

\usepackage{algorithm}
\usepackage{algpseudocode}
\usepackage{booktabs}
\usepackage{multirow}
\usepackage{wrapfig}
\usepackage{listings}
\usepackage{siunitx}
\usepackage{pgfplotstable}
\usepackage{array}
\usepackage{subcaption}
\usepackage{newfloat}

\usepackage{pdflscape}
\usepackage{xspace}

\usepackage{diagbox} 

\usepackage[]{todonotes}

\newcommand*{\etal}{\emph{et~al.}\@\xspace}
\newcommand\YAMLcolonstyle{\color{red}\mdseries}
\newcommand\YAMLkeystyle{\color{black}\bfseries}
\newcommand\YAMLvaluestyle{\color{blue}\mdseries}

\makeatletter

\newcommand\language@yaml{yaml}

\expandafter\expandafter\expandafter\lstdefinelanguage
\expandafter{\language@yaml}
{
  keywords={true,false,null,y,n},
  keywordstyle=\color{darkgray}\bfseries,
  basicstyle=\YAMLkeystyle,                                 
  sensitive=false,
  comment=[l]{\#},
  morecomment=[s]{/*}{*/},
  commentstyle=\color{purple}\ttfamily,
  stringstyle=\YAMLvaluestyle\ttfamily,
  moredelim=[l][\color{orange}]{\&},
  moredelim=[l][\color{magenta}]{*},
  moredelim=**[il][\YAMLcolonstyle{:}\YAMLvaluestyle]{:},   
  morestring=[b]',
  morestring=[b]",
  literate =    {---}{{\ProcessThreeDashes}}3
                {>}{{\textcolor{red}\textgreater}}1     
                {|}{{\textcolor{red}\textbar}}1 
                {\ -\ }{{\mdseries\ -\ }}3,
}

\lst@AddToHook{EveryLine}{\ifx\lst@language\language@yaml\YAMLkeystyle\fi}
\makeatother

\newcommand\ProcessThreeDashes{\llap{\color{cyan}\mdseries-{-}-}}

\usepackage{wasysym}
\newcommand{\use}{Use}
\newcommand{\useBy}{UsedBy}

\DeclareFloatingEnvironment[
    fileext=los,
    listname={List of Listings},
    name=Listing,
    placement=!h,
    within=none,
]{listing}

\lstdefinestyle{docker}{
  language=bash,
  morekeywords={RUN,FROM,MAINTAINER},
  showstringspaces=false,
  frame=none,
}
\lstdefinelanguage{ansible}{
  morekeywords={name,vars,hosts,tasks,roles,role},
  keywordstyle=\bfseries,
  morecomment=[l][\textit]\#,
  morecomment=[s][\bfseries]{\{\{}{\}\}},
}
\lstdefinestyle{ansible}{
  language=ansible,
  basicstyle=\scriptsize\ttfamily,
}

\def\postbreak{%
  \raisebox{0ex}[0ex][0ex]{\ensuremath{\hookrightarrow\space}}}

\lstdefinestyle{searchstringstyle}{
	basicstyle=\ttfamily\footnotesize,
	breakatwhitespace=false,         
	breaklines=true,                 
	captionpos=t,                    
	keepspaces=true,                 
	numbers=none,                    
	numbersep=5pt,                  
	showspaces=false,                
	showstringspaces=false,
	showtabs=false,                  
	tabsize=2,
	frame=single
}

\definecolor{mauve}{rgb}{0.58,0,0.82}
\definecolor{dkgreen}{rgb}{0,0.6,0}
\definecolor{gray}{rgb}{0.5,0.5,0.5}

\makeindex             

\begin{document}

\frontmatter




\mainmatter




\title{Promises and Perils of Mining \\ Software Package Ecosystem Data}
\author{Raula Gaikovina Kula \and Katsuro Inoue \and Christoph Treude}
\institute{Raula Gaikovina Kula \at Nara Institute of Science and Technology, Japan, \email{raula-k@naist.jp}
\and Katsuro Inoue \at Nanzan University, Japan \email{ inoue599@nanzan-u.ac.jp}
\and  Christoph Treude \at The University of Melbourne, Australia \email{christoph.treude@unimelb.edu.au}}

\maketitle
\label{PPM:ch}
\abstract*{The use of third-party packages is becoming increasingly popular and has led to the emergence of large software package ecosystems with a maze of inter-dependencies. Since the reliance on these ecosystems enables developers to reduce development effort and increase productivity, it has attracted the interest of researchers: understanding the infrastructure and dynamics of package ecosystems has given rise to approaches for better code reuse, automated updates, and the avoidance of vulnerabilities, to name a few examples. But the reality of these ecosystems also poses challenges to software engineering researchers, such as: How do we obtain the complete network of dependencies along with the corresponding versioning information? What are the boundaries of these package ecosystem? How do we consistently detect dependencies that are declared but not used? How do we consistently identify developers within a package ecosystem? How much of the ecosystem do we need to understand to analyse a single component? How well do our approaches generalise across different programming languages and package ecosystems? In this chapter, we review promises and perils of mining the rich data related to software package ecosystems available to software engineering researchers.}

\abstract{The use of third-party packages is becoming increasingly popular and has led to the emergence of large software package ecosystems with a maze of inter-dependencies. Since the reliance on these ecosystems enables developers to reduce development effort and increase productivity, it has attracted the interest of researchers: understanding the infrastructure and dynamics of package ecosystems has given rise to approaches for better code reuse, automated updates, and the avoidance of vulnerabilities, to name a few examples. But the reality of these ecosystems also poses challenges to software engineering researchers, such as: How do we obtain the complete network of dependencies along with the corresponding versioning information? What are the boundaries of these package ecosystems? How do we consistently detect dependencies that are declared but not used? How do we consistently identify developers within a package ecosystem? How much of the ecosystem do we need to understand to analyse a single component? How well do our approaches generalise across different programming languages and package ecosystems? In this chapter, we review promises and perils of mining the rich data related to software package ecosystems available to software engineering researchers.}


\section{Introduction}
\label{PPM:sec:definition}

Third-party libraries are a great way for developers to incorporate code without having to write their own for every functionality required. By using these libraries, developers can save time and energy while still getting the functions they need.
Using third-party libraries is becoming increasingly popular and has led to the emergence of large software package ecosystems such as npm. While these ecosystems offer many benefits, they also come with risks, such as software vulnerability attacks \cite{Chinthanet:ASE2020}.

Large software package ecosystems are a treasure trove for researchers who can investigate a wide range of questions. For example, by studying activity in large ecosystems, researchers can identify which libraries are the most popular and learn what characteristics make them successful \cite{kikas.2017,decan:emse:2019}.
Additionally, research on large ecosystems can help developers understand how to protect their code from malicious actors who may attempt to exploit vulnerabilities or insert malware into popular libraries.
Studying large software package ecosystems can help us better understand the dynamics of open source development in general. Open source development is a complex process that involves many different stakeholders working together (or sometimes competing) to create valuable code that anyone can use or improve upon. By understanding how these interactions play out in different types of ecosystem structures -- including those with many small projects versus few very large ones -- we can develop insights that might be applicable more broadly across other types of collaborative systems.

In this chapter, we identify and discuss promises and perils during the mining process, ranging from planning what information to mine from the ecosystem to analysing and visualising the mined data. 
Therefore, the chapter is broken down into these logical processes of mining ecosystem data: 1) Planning what Information to Mine, 2) Defining Components and their Dependencies, 3) Defining Boundaries and Completeness, and 4) Analysing and Visualising the Data.

This chapter is intended for researchers and practitioners who are interested in exploring and exploiting software package ecosystem information from a diverse range of sources that are publicly available. 
We also highlight the pitfalls to consider during the mining process, particularly when these pitfalls could lead to a misinterpretation of the analysis and results. 
The chapter is written in a manner that encourages newcomers who have little or no experience or who are interested in utilising ecosystem data across different disciplines outside of software engineering.
Our goal is to get new researchers quickly accustomed to gathering ecosystem information for their research.

\section{A Component-based Software Ecosystem}

Defined as a component-based software ecosystem, we suggest using the term `software package ecosystem' as a suitable term for the symbiotic relationships among third-party library components (as software projects or repositories), as these libraries and their dependent clients coexist on the same technological platform, therefore sharing the same environment and other internal and external factors (e.g., security threats, sharing contributions, etc.).
Please refer to the Introduction chapter for an in-depth definition of the different types of software ecosystems.
We present our interpretation of the software package ecosystem in Kula et al.~\cite{KulaSANER18}, where we formally define a package ecosystem using a Software Universe Graph (SUG).
This is modelled as a structured abstraction of the evolution of software systems and their library dependencies over time.

\begin{figure*}
	\centering
	\includegraphics[width=.7\textwidth]{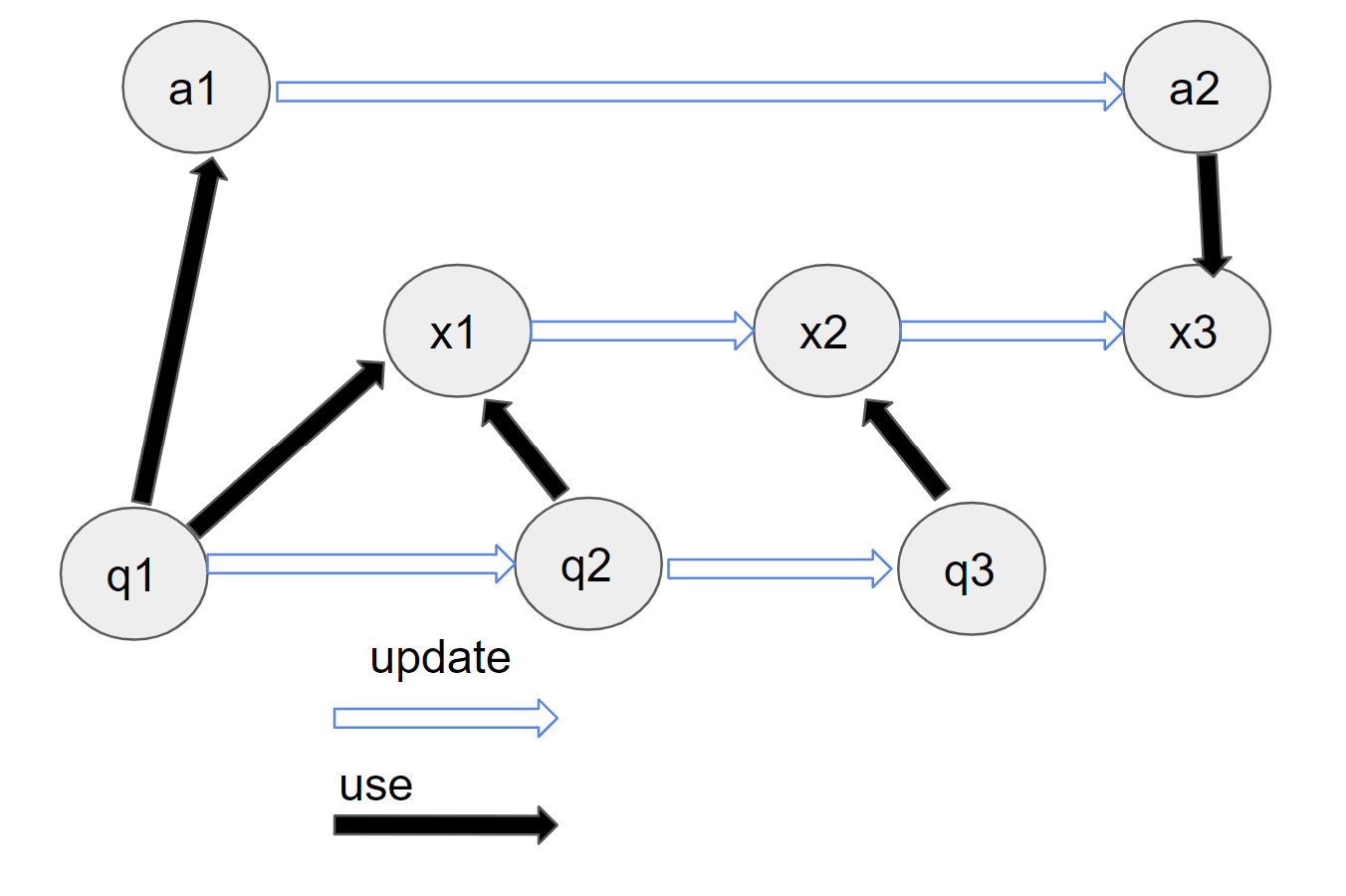}
	\caption{Conceptual example of the Software Universe Graph, depicting the use and update relationships between different software units.}
	\label{fig:SUG}
\end{figure*}

\paragraph{\textbf{Component-based Representation as a Software Universe Graph}}
\label{PPM:sec:SUG}

First introduced by Kula et al.~\cite{KulaSANER18}, the \textit{Software Universe Graph} (SUG) is a structural abstraction of the software ecosystem of third-party libraries.
Figure \ref{fig:SUG} provides an illustration of the different relationships within the graph.
Let $G= (N,E)$ represent a graph $G$. $N$ is a set of nodes, each node representing a software unit. 
We define a software unit as a version instance of any software program. 

The authors then present the \textit{use} and \textit{update} relationships that exist in the ecosystem.
Hence, the edges $E$ are composed of $E_{use}$ and $E_{update}$. $E_{use}$ is a set of \textit{use-relations} and $E_{update}$ is a set of \textit{update-relations}.

\begin{definition}
An edge $u \rightarrow v \in E_{use}$ means that $u$ uses $v$. The defined functions of $E_{use}$ are:

\begin{equation}
\small
\small \use(u)\equiv \{v|u \rightarrow v\}
\normalsize
\end{equation}
\begin{equation}
\small
\small \useBy(u)\equiv \{v|v \rightarrow u\}
\normalsize
\end{equation}
\end{definition}

Use-relations can be extracted from either the source code or configuration files. 
As shown in Figure \ref{fig:SUG}, node $a1$ uses node $x1$. 
In addition, node $x1$ is used by nodes $a1$, $q1$, and $q2$. Parallel edges for node pairs are not allowed.

\begin{definition}
We represent an update relation from node $a$ to $b$ using $ a \Rightarrow b $, which means that the newer update $b$ was released from node $a$ and is defined as:
\begin{equation}
\small a \Rightarrow b \in E_{update}
\end{equation}
\end{definition}

Update relations refer to when a successive release of a software unit is made available. Figure \ref{fig:SUG} shows that node $q1$ is first updated to node $q2$. Later, node $q2$ is updated to the latest node $q3$. Hence, $q1 \Rightarrow q2 \Rightarrow q3$.
Note that an update should not be confused with forking. 
We distinguish a fork as a separate software unit. 
Each node in the SUG should be denoted by three attributes: \texttt{<name,release,time>}.  
For a node $u$, we define:

\begin{itemize}
	\item \textbf{u.name} Name is the string representation identifier of a software unit.
	We introduce the name axiom: For nodes $u$ and $v$, if $u \Rightarrow v$, then $u.name = v.name$ holds.
	
	\item \textbf{u.release}. Release refers to the specific assigned change reference for a software unit. For nodes $u$ and $v$, if $u \Rightarrow v$
	then $v$ is the immediate successor of $u$. Note that the versioning pattern may vary from project to project. 
	\item \textbf{u.time}. Time refers to the time stamp at which node $u$ was released. For nodes $u$ and $v$ of $u \Rightarrow v$, $u.time < v.time$.
\end{itemize}

\begin{figure}
	\centering
	\includegraphics[width=.8\textwidth]{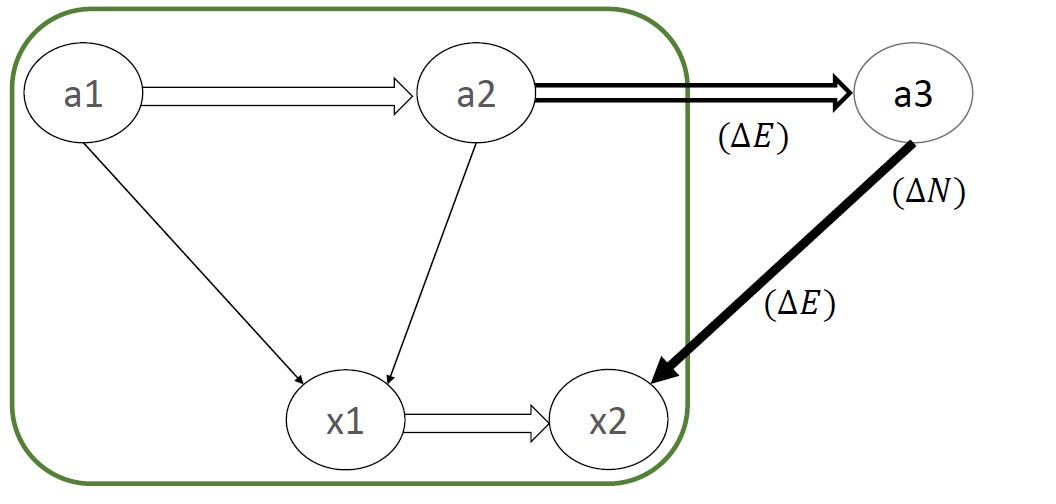}
	\caption{Temporal property of the SUG}
	\label{fig:SUGTemp}
\end{figure}

\begin{definition}
    
	The SUG has temporal properties.
This describes the simultaneity or ordering in reference to time. Let SUG $G = (N, E) $ be at time $t$. At time $t^{\prime} > t$, we observe an extension of $G$, such that:

\begin{equation}
\small G^{\prime} = (N \cup \Delta N, E \cup \Delta E)
\end{equation}
where $\Delta E \cap (N \times N) = \emptyset$
\end{definition}

Figure \ref{fig:SUGTemp} illustrates the temporal properties of the SUG. 
Here, it is observed that $G'$ is composed of $G$ augmented with newly added node $a3$ and its corresponding $a3 \rightarrow x2$ and $a2 \Rightarrow a3$ relations.
A SUG grows monotonically over time with only additions.
Here, we consider that modification or deletion changes on the SUG do not occur. 

\begin{definition}
    A timed SUG specifies the state of the SUG at any point in time.
So for an SUG $G=(N,E)$, we represent a timed SUG $G_{t}$ at time $t$ as a sub-graph of $G$. Formally,
\begin{equation}
\small G_t\equiv(N_{t}, E_{t})
\end{equation}
where $N_{t} = \{u|u \in N, u.time \leq t \}$ and $E_t = \{ e | e \in E \wedge e \in  N_t \}$
\end{definition}

\section{Data Sources}
Researchers can use various datasets to model the ecosystem using the SUG model of usage and update relationships.
The most obvious data source that has revolutionised data mining in the software engineering domain is the GitHub platform. 
Established in 2008, and then purchased by Microsoft in 2020, GitHub is home to various popular Open Source Software. 
GitHub is built on the git version control system and is useful for storing all changes made to a repository. 
In the case of the SUG, a GitHub repository can represent one software unit, whose depend relations can be extracted via a configuration file (such as the package.json file for JavaScript projects).
The repository should also contain the release information that holds the update relations.
Due to its large size, researchers and the GitHub team have made available datasets for researchers to mine, for example through the GitHub API/Graph QL.\footnote{\url{https://docs.github.com/en/graphql}} This is the backend Application Programming Interface (API) that can be used to query large amounts of data on GitHub. Most researchers use the API to download and mine information from the GitHub platform. 
It is important to note that while GitHub introduced a new feature of Dependency Graphs to map the depend relationship,\footnote{\url{https://docs.github.com/en/code-security/supply-chain-security/understanding-your-software-supply-chain/about-the-dependency-graph}} most older projects do not have this feature.
In this case, the researcher would need to manually extract and query the configuration files for dependency information. 

We refer to the first chapter for additional information on data sources for mining software ecosystems. 

\section{Promises and Perils}
\label{PPM:sec:promisesperils}

Using the SUG model of depend and use relations and the available datasets, we can now present our promises and perils of mining ecosystem information.

\subsection{Planning What Information to Mine}

\textbf{Promise 1.}\textit{
Researchers can access and link heterogeneous data related to software package ecosystems, e.g., package registries and bug trackers.}\\

When planning what information to mine from the ecosystem, researchers do not need to limit themselves to the usage and update relationship information.
Platforms that host software repositories include other software management systems such as bug trackers.
For example, GitHub allows researchers to manage GitHub Pull Requests, Issues, and Discussions not only for one project, but for multiple projects.
GitHub provides three management systems that are related to a software repository:

\begin{itemize}
    \item \textit{GitHub Discussions}\footnote{\url{https://docs.github.com/en/discussions}} - The GitHub Discussions forum is a collaborative communication forum for the community around an open source or internal project. Community members can ask and answer questions, share updates, have open-ended conversations, and follow along on decisions affecting the community's way of working.
    \item \textit{GitHub Pull Requests}\footnote{\url{https://docs.github.com/en/pull-requests}} - Pull Requests allow other developers from an ecosystem to make a contribution to a software repository. Pull requests also allow maintainers to discuss and review potential changes with collaborators and add follow-up commits before changes are merged into the software.
    \item \textit{GitHub Issues}\footnote{\url{https://docs.github.com/en/issues}} - Issues are used to track ideas, feedback, tasks, or bugs for work on GitHub.
\end{itemize}

These three systems are examples of how developers contribute to both their own and other projects. 
Hence, to incorporate this information, we can extend the SUG model, creating a model that includes a contribution relationship \cite{wattanakriengkrai2022giving}.


\begin{definition}
	A Dependency-Contribution graph incorporates contributions by developers whose libraries are involved in dependency relationships. 
\end{definition}

In this work \cite{wattanakriengkrai2022giving}, the authors explore the congruence between dependency updates and developer contributions, based on the original concept of social-technical congruence \cite{stcCataldo2008} where developers contribution patterns are congruent with their coordination needs. Hence, the goal is to identify contributions that are congruent to dependency updates.
As shown in Figure \ref{fig:lib} the authors extend from the typical SUG graph model where $lib_i$ depends (use) on  $lib_k$ and  $lib_j$, while  $lib_j$ also depends on $lib_k$, to the example shown in Figure \ref{fig:dc-graph}.
Different to the SUG, the graph captures developers and their contributions (i.e., the square as $dev_x$ and $dev_y$ represent two different developers making a contribution).
Here contributions are defined as $c$ (Pull Request or Issue) that were submitted to both a library and the client that depends on that library.
Hence, the graph can show contributions that are congruent to dependency changes for a software unit. 

\begin{figure}[t]
     \centering
     \begin{tikzpicture}[
         roundnode/.style={circle, fill=black, minimum size=5mm},
        squarenode/.style={fill=black, text=red, minimum size=5mm},
     ]
    \begin{scope}
         \node[roundnode, label=above:$lib_i$] (s2_proji) at (3, 2.5) {};
        \node[roundnode, label=below:$lib_j$] (s2_projj) at (4,0) {};
         \node[roundnode, label=below:$lib_k$] (s2_projk) at (2,0) {};
     \end{scope}

     \begin{scope} [every edge/.style={draw=gray, very thick}]
         \path [->] (s2_proji) edge (s2_projj);
         \path [->] (s2_proji) edge (s2_projk);
         \path [->] (s2_projj) edge (s2_projk);
     \end{scope}
     \end{tikzpicture}
    
     \caption{Example dependency graph for a given time period}
     \label{fig:lib}
     \vspace{2ex}
 \end{figure}
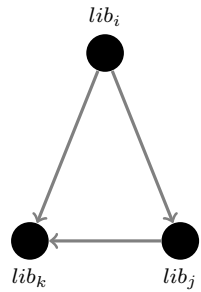
\begin{figure}[t]
    \centering

    \begin{tikzpicture}[
        roundnode/.style={circle, fill=black, minimum size=5mm},
        squarenode/.style={fill=black, text=red, minimum size=5mm},
    ]
    \begin{scope}
        
        \node[roundnode, label=right:$lib_i$] (s2_proji) at (4, 1.5) {};
        \node[roundnode, label=below:$lib_j$] (s2_projj) at (5,0) {};
        \node[roundnode, label=below:$lib_k$] (s2_projk) at (3,0) {};
        
        \node[squarenode, label=below:$dev_x$] (s3_devx) at (2, 3.5) {};
        
        \node[squarenode, label=below:$dev_y$] (s3_devy) at (6, 3.5) {};
        
    \end{scope}

    \begin{scope} [every edge/.style={draw=gray, very thick}]
        \path [->] (s2_proji)  edge  (s2_projj);
        \path [->] (s2_proji) edge (s2_projk);
        \path [->] (s2_projj) edge (s2_projk);
        
    \end{scope}
    \begin{scope} [every edge/.style={draw=gray, thick, double distance=2pt}]
        \path [->] (6, 2.8) edge node[left = 2mm] {$contribute$} (s2_proji);
        \path [->] (6, 2.8) edge[bend left=15] node[right = 1mm] {$contribute$} (s2_projj);
        \path [->] (2, 2.8) edge node[right = 2mm] {$$} (s2_proji);
        \path [->] (2, 2.8) edge[bend right=15] node[left = 1mm] {$contribute$} (s2_projk);
    \end{scope}
    \end{tikzpicture}
\end{figure}
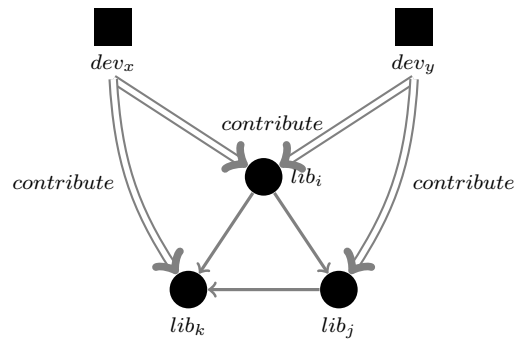
\begin{figure}[t]
    \centering
    \begin{tikzpicture}[
        roundnode/.style={circle, fill=black, minimum size=5mm},
        squarenode/.style={fill=black, text=red, minimum size=5mm},
    ]
    \end{tikzpicture}
\caption{Example Dependency-Contribution graph showing relationships between contributions and dependencies}
 \label{fig:dc-graph}
\end{figure}

This is just one example of the type of research that is enabled by access to heterogeneous data related to software package ecosystems.

\smallskip\noindent\textbf{Peril 1.}\textit{
 Developers might use different identifiers when contributing to different parts of a software package ecosystem, e.g., when contributing to different libraries.}\\
 
When modelling using such graphs, there is a threat that contributors may use multiple identifiers (i.e., $c_x$ and $c_y$ are the same contributor).
This is a well-known research problem, and there has been work to merge these accounts, such as \cite{wiese2016mailing}.
GitHub has introduced mechanisms such as two-factor authentication\footnote{\url{https://docs.github.com/en/authentication/securing-your-account-with-two-factor-authentication-2fa/configuring-two-factor-authentication}} to counteract the issue of multiple identifiers.
This is since developers might be less likely to switch accounts if it requires cumbersome authentication.

\smallskip\noindent\textbf{Peril 2.}\textit{
Developers' contributions to software package ecosystems might be interspersed with bot contributions, e.g., automated dependency updates.}\\

The rise of automation and artificial intelligence has led to much work on the integration of automated scheduling (i.e., bots) into software development workflows \cite{Storey2016, Farooq2016, Wessel2018, Erlenhov2019, bot_modify_wf} to name a few. These bots are designed to perform specific tasks within a software package ecosystem. For example, a bot may be programmed to automatically update dependencies, test code changes, or deploy software to production. As an example, the Google APIs repo-automation-bots project lists bots for automated labelling of issues and pull requests, automated approval of pull requests, and triggering releases.\footnote{\url{https://github.com/googleapis/repo-automation-bots}}
Bots perform common maintenance tasks in many software projects and are now commonplace \cite{Beschastnikh2017, Urli2018,BIMAN,bot_or_not}.
Especially with bots such as dependabot (automated pull requests to update configurations to reduce the risk of vulnerability threats),\footnote{\url{https://github.com/dependabot}} more and more automation has caused a lot of noise in the contributions between projects.
There are also bots for communication and documentation \cite{Urli2018, Lin2016, Lebeuf2017a}.

To be able to draw accurate conclusions about what humans are doing in software package ecosystems, researchers should consider distinguishing between bot and human contributions.
It is also important to differentiate this from other contributions \cite{maeprasart2022understanding}.
The research community has responded well, with a wide range of techniques and tools to mitigate this peril \cite{Bodegha2021, golzadeh2022accuracy}.

\smallskip\noindent\textbf{Peril 3.}\textit{
Not all developer activities in software package ecosystems are accessible to the public, e.g., library use in proprietary settings.
}\\

Not all developer activities in software package ecosystems are accessible to the public, e.g., when the boundary between open source and industry is blurred \cite{stol2014inner}, which presents a challenge for researchers who aim to study the development process. This is particularly true in proprietary settings where software development is performed behind closed doors or is open source for a limited time period, thus resulting in the artefacts not permanently being publicly available.
This can make it difficult to understand the broader ecosystem in which a software project is developed.
Proprietary settings may lead to non-standardisation in software development practises. Different software projects may use different management systems and tools, making it difficult to accurately compare and analyse software development activities across various projects. For example, some projects may use communication, documentation, and other management tools not captured on the same platform \cite{montgomery2022alternative}. For example, some projects might use Bugzilla instead of issues and pull requests for their bug and code review systems, while others may use Discord, Slack channels, or email threads for their communication needs.

This lack of standardisation in software development practises presents a challenge for researchers who study the software package ecosystem and understand the development process. To address this issue, researchers should strive to collect data from a diverse set of projects to gain a comprehensive understanding of the software package ecosystem. In addition, researchers may need to adjust their methodologies or data collection techniques to accommodate the different tools and practises used by different software projects.

\subsection{Defining Components and their Dependencies}

\smallskip\noindent\textbf{Promise 2.}\textit{
Researchers can access a software package ecosystem's dependency network through package managers and registries, e.g., npm lists the dependencies and dependents for over a million libraries.}\\

With the rise of curated datasets like libraries.io, researchers can now recover and model dependency relations between software units using pre-extracted datasets.
Table \ref{tab:PM_features} shows examples of popular package managers mined from the libraries.io dataset in 2020. 

\begin{table*}[]
\caption{Summary of 13 package managers from libraries.io as ranked by TIOBE in 2020}
 \label{tab:PM_features}
 \centering
\begin{tabular}{@{}llrlll@{}}
\toprule
\begin{tabular}[c]{@{}l@{}}Package \\ Ecosystem\end{tabular} & \begin{tabular}[c]{@{}l@{}}Programming \\ Language\end{tabular} &  \begin{tabular}[c]{@{}l@{}}Tiobe \\ Rank\end{tabular} & Environment & \begin{tabular}[c]{@{}l@{}}Dependency \\ Tree\end{tabular} & \begin{tabular}[c]{@{}l@{}}Package   \\Archive link\end{tabular}\\ \midrule
PyPI & Python & 2 & Python & Flat & pypi.org  \\
Maven & Java & 3 & JVM & Flat & Maven.org  \\
Bower & JavaScript & 7 & Node.js & Flat & bower.io  \\
Meteor & JavaScript & 7 & Node.js & Nested & atmospherejs.com \\
npm & JavaScript & 7 & Node.js & Nested (v2) & npmjs.com  \\
Packagist & PHP & 8 & PHP & Flat & packagist.org \\
Puppet & Ruby & 13 & Ruby MRI & Flat & forge.puppet.com  \\
RubyGems & Ruby & 13 & Ruby MRI & Flat & rubygems.org  \\
CRAN & R & 14 & RStudio & Flat & cran.r-project.org  \\
CPAN & Perl & 15 & Perl & Flat & metacpan.org  \\
GO & Golang & 20 & Go & Flat & pkg.go.dev  \\
NuGet & C\#, VB & 5, 6 & .NET & Flat & nuget.org \\
Anaconda & Python, R, C\# & 2, 14, 5 & Anaconda & Flat & anaconda.org  \\ \bottomrule
\end{tabular}
\end{table*}

\smallskip\noindent\textbf{Peril 4.}\textit{
Different software package ecosystems define the concept of ``dependency'' differently, e.g., by allowing or not allowing different versions of a library on the same dependency tree.
}\\

Different software package ecosystems have varying definitions of what constitutes a dependency. For example, some ecosystems may allow multiple versions of a library to exist on the same dependency tree, while others may restrict developers to a single version of a library \cite{Islam}. These restrictions are often based on the programming language being used, as different languages have different approaches to managing dependencies. It is important to consider the restrictions on dependency relationships when studying software package ecosystems, as they can have a major impact on the development process. For example, the ability to use multiple versions of a library on the same dependency tree can greatly simplify the process of updating dependencies and can make it easier to resolve conflicts between libraries.

\begin{figure}
	\centering
	\includegraphics[width=.8\textwidth]{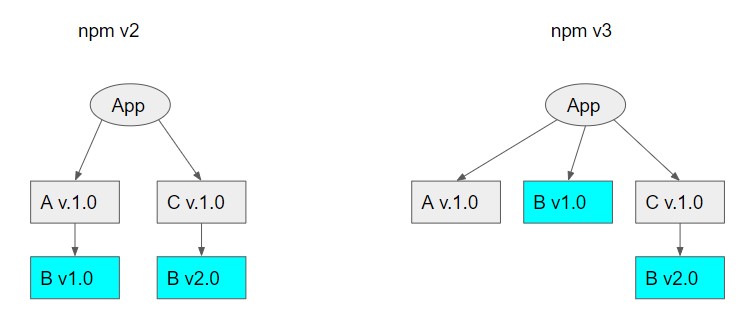}
	\caption{Difference between flat and nested dependencies}
	\label{fig:nest}
\end{figure}

One way to visualise the impact of these restrictions is to compare the difference between a nested dependency tree and a directed dependency tree, as shown in Figure \ref{fig:nest}.\footnote{Taken from \url{https://npm.github.io/how-npm-works-docs/npm3/how-npm3-works.html}} This distinction is important because it highlights the different ways that a software unit can depend on different versions of the same library.
In this example, npm v3 creates the dependency tree based on the installation order, therefore flattening unnecessary nested dependencies (i.e., B v1.0 in cyan). This reduces the complexity of a nested tree by resolving some of the transitive dependencies (nested dependencies).

\smallskip\noindent\textbf{Peril 5.}\textit{
Developers might declare a dependency to other parts of a software package ecosystem but not use it, e.g., because they failed to update its removal.
}\\

It is common for developers to declare dependencies on other parts of the software package ecosystem but not always use them. This can happen for various reasons, such as forgetting to remove the dependency after it is no longer needed. This can pose a challenge for researchers who are trying to extract dependencies from package managers, like those in configuration files, as there may be inconsistencies between the listed dependencies and what is actually being compiled and used by the code. This can lead to a biased understanding of the software package ecosystem and the relationships between software components.

To address this issue, there have been numerous efforts to track the actual library dependencies compiled and executed in software systems. These efforts aim to provide a more accurate understanding of the dependencies and the relationships between software components. For example, research has been conducted on the use of dynamic analysis to track compiled dependencies in real time and on the development of tools to automatically detect and track executed dependencies \cite{Zapata:ICSME2018, Ponta2018,Chinthanet:ASE2020}.

\subsection{Defining Boundaries and Completeness}

\smallskip\noindent\textbf{Promise 3.}\textit{
Researchers can use the boundaries of software package ecosystems to study communities of developers, e.g., developers contributing to and/or benefiting from the npm ecosystem.
}\\

Following Promise 2, the emergence of package managers has also led to studies that approximate software communities.
Using the libraries.io dataset, researchers were able to study projects that host libraries that use package managers.
Researchers have used this dataset to compare different library ecosystems \cite{kikas.2017,decan:emse:2019,CogoDown2019}.

\smallskip\noindent\textbf{Peril 6.}\textit{
Package managers do not always represent software package ecosystems, their communities, or their sub-communities, e.g., in cases where multiple package managers exist.
}\\

Package managers are a fundamental aspect of software package ecosystems, but do not always fully represent the complex relationships and interactions that occur within a community of developers and users, as shown in Table~\ref{tab:PM_features}. In some cases, multiple package managers exist for the same programming language, creating a complex landscape of software libraries and dependencies that are not always easily understood. For instance, Bower and Meteor manage npm libraries, which can lead to confusion and overlap in the management of dependencies.

Similarly, Java, Scala, Android, and other Java-based open source communities all use the Maven package manager, but each of these communities has its own unique set of libraries, dependencies, and development practises. Researchers should be aware of the limitations of package managers when studying software package ecosystems, and consider the broader context and relationships that exist within these communities. 

\smallskip\noindent\textbf{Peril 7.}\textit{
Lack of activity in parts of a software package ecosystem does not necessarily indicate project failure, e.g., when highly depended-upon libraries are feature-complete.
}\\

It is important to note that lack of activity in a part of a software package ecosystem does not always mean project failure \cite{coelho2017modern}. In some cases, highly relied-upon libraries that have reached feature-completeness may see little activity, but continue to be used by the software community. 

However, it is still important to consider the long-term sustainability of these libraries, especially given the rate at which technology and software development practises change. This has become a topic of interest in recent years, and researchers have explored best practises for sustaining open source projects and ensuring their continued success \cite{Ait2022,valiev2018ecosystem}. Understanding the factors that contribute to project sustainability is important to ensure the longevity and continued growth of software package ecosystems.

\smallskip\noindent\textbf{Peril 8.}\textit{
Sampling from a software package ecosystem is challenging since sub-setting might alter the dependency network, e.g., by breaking dependency chains.
}\\

Sampling from a package ecosystem is not straightforward, as the sample composition can be significantly affected due to missing dependency links between libraries. For instance, a subset of the ecosystem might alter the dependencies between libraries, leading to the breakdown of the dependency chains. This could lead to an incomplete picture of the software package ecosystem, leading to incorrect conclusions from a study. To minimise this risk, researchers should carefully consider the boundaries of their study and choose the appropriate sampling method based on the research questions and goals. For example, researchers could focus on popular, highly dependent, or risk-vulnerable aspects of the ecosystem as a starting point. 
For some ecosystems, the number of downloads, GitHub stars, and watchers are other aspects for the researcher to utilise.

\smallskip\noindent\textbf{Peril 9.}\textit{
Sampling from a software package ecosystem is challenging since the dependency network changes over time, e.g., when dependencies are added, removed, upgraded, or downgraded.
}\\

The dynamic nature of package ecosystems and the constant changes to their dependencies can impact the generalisability of the results. Therefore, it is important to also consider the time granularity of the analysis. For example, if the goal is to understand the evolution of dependencies over time, a finer time granularity may be necessary to capture the smaller changes and trends. However, if the goal is to understand the overall structure and relationships within the ecosystem, a coarser time granularity may be sufficient. Based on recent studies \cite{wattanakriengkrai2022giving,valiev2018ecosystem,Mirsaeedi:icse2020, Brindescu:emse2020, Nassif:icsme2017}, a three-month window seems appropriate for some studies.
Another level of granularity to consider is the size of the component. For instance, there are cases where a single package may contain more than one repository, especially for large library frameworks. 
The granularity also depends on the nature of the ecosystem itself. For instance, researchers should understand whether the ecosystem comprises library packages (e.g., PyPI), plugins (e.g., Eclipse), or is a library distribution (e.g., Android).

\subsection{Analysing and Visualising the Data}

\textbf{Peril 10.}\textit{
Analysing and visualising entire software package ecosystems is challenging due to their size, e.g., in terms of nodes and edges in the network.
}\\

The size of software package ecosystems implies large data sets, which can be overwhelming for tools and algorithms to analyse and display. Therefore, it may be necessary to make choices about the granularity of the data included in the analysis and visualisation. Another alternative is to focus on the most critical parts of the software package ecosystem, such as the high-level structure, highly dependent packages, or parts of the system that pose a risk to security and reliability. 
The key is to strike a balance between detail and simplicity, providing a meaningful representation of the ecosystem while being able to handle the complexity of its size.

\section{Application: When to Apply Which Peril}
\label{PPM:sec:application}

We include a disclaimer stating that not all perils are applicable to every mining situation. To demonstrate the practical application of our perils and their mitigation, we present two case studies that involve mining the software package ecosystem. Each case study has a distinct research objective and focusses on a specific dataset to be mined.

\subsection{Two Case Studies}
Table \ref{tab:cases} presents the two case studies we have selected for this analysis.
The \textit{first case} involves mining for contributions congruent to dependency updates \cite{wattanakriengkrai2022giving}. 
In this work, the authors mine GitHub repositories for Pull Requests and Issues that were submitted and merged congruent to dependency updates within the npm ecosystem. 
The \textit{second case} involves mining communication data for the Eclipse ecosystem \cite{Nugroho2021}. Although the second case does not mine for dependency relations (i.e., use relations),  we show that these perils still apply when mining for other relationships in an ecosystem.
Moreover, the second case studies the Eclipse ecosystem, which is a different dataset compared to the more popular GitHub dataset.

\subsection{Applying Perils and their Mitigation Strategies}
Table \ref{tab:perilsapp} provides a summary of the perils that can be applied to each of the case studies. We will now go into the details of mitigation strategies based on these perils. 
For better organisation and understanding, we have grouped the perils according to the four logical processes for mining.

\smallskip\noindent\textbf{Information to Mine}. 
The first set of mitigation strategies, which addresses perils 1-3, focusses on planning which information to mine. There are two primary strategies that researchers can employ:

\begin{enumerate}
    \item Researchers should use research tools and techniques to remove noise and other biases in the dataset, such as bot detection and the handling of multiple identities. This strategy was implemented in both case studies, as contributions and discussions often have the potential to involve bots or developers with multiple identities.
\item Depending on the research goals, researchers should recognise that not all contributions are equal and filter the dataset accordingly.
\end{enumerate}

We applied these two strategies to both cases. In the first case, the goal was to capture all congruent contributions, so we filtered out contributions made to libraries without dependencies. Since all npm packages are listed in the registry, Peril 3 (private activities) did not apply.
In the second case, we addressed Peril 1 by conducting a qualitative analysis to ensure that the member identities were not duplicated, as Eclipse developers were known to change identities. To mitigate Peril 2, we removed bot responses. For the second case, since all forum data is made public, Peril 3 did not apply.

\begin{table}
\centering
\caption{Description of the research objectives and datasets for the case studies}
 \label{tab:cases}
\begin{tabular}{lp{5cm}r} 
\toprule
\textbf{Case Study} & \textbf{Research Objective}                                                          & \textbf{Datasets}           \\
\midrule
Wattanakriengkrai \etal \cite{wattanakriengkrai2022giving}             & Explore code contributions between library and client (i.e, use-relations)  & libraries.io\\
& &  GitHub API \\
Nugroho \etal \cite{Nugroho2021}                & Explore discussion contributions between contributors (i.e., contributions) & Eclipse API                \\
\bottomrule
\end{tabular}
\end{table}

\begin{table}
\centering
\caption{Application of each peril to the case studies}
 \label{tab:perilsapp}
\begin{tabular}{rp{8cm}ccc} 
\toprule                   
& \textbf{Perils}       & case 1      & case 2                 \\
                                                                               & & npm  & Eclipse   \\
\midrule
\textbf{P1} &Developers might use different identifiers when contributing to different parts of a software package ecosystem, e.g., when contributing to different libraries.                             &    \CheckedBox         &          \CheckedBox           \\ 
\textbf{P2} & Developers' contributions to software package ecosystems might be interspersed with bot contributions, e.g., automated dependency updates.                                                      &      \CheckedBox               &      \CheckedBox               \\ 
\textbf{P3} & Not all developer activities in software package ecosystems are accessible to the public, e.g., library use in proprietary settings.                                                         &         -           &      \CheckedBox               \\ 
\textbf{P4} & Different software package ecosystems define the concept of \`{}\`{}dependency'' differently, e.g., by allowing or not allowing different versions of a library on the same dependency tree. &         \CheckedBox       &     -                            \\ 
\textbf{P5} & Developers might declare a dependency to other parts of a software package ecosystem but not use it, e.g., because they failed to update its removal.                                        &     -           &  -    \\
\textbf{P6} &Package managers do not always represent software package ecosystems, their communities, or their sub-communities, e.g., in cases where multiple package managers exist.                     &     -                   &  -                  \\
\textbf{P7} & Lack of activity in parts of a software package ecosystem does not necessarily indicate project failure, e.g., when highly depended-upon libraries are feature-complete.                     &       \CheckedBox         &  -                           \\
\textbf{P8} &Sampling from a software package ecosystem is challenging since sub-setting might alter the dependency network, e.g., by breaking dependency chains.                                         &        \CheckedBox        &     \CheckedBox                                \\
\textbf{P9} &Sampling from a software package ecosystem is challenging since the dependency network changes over time, e.g., when dependencies are added, removed, upgraded, or downgraded.               &       \CheckedBox         &     -                          \\
\textbf{P10} &Analysing and visualising entire software package ecosystems is challenging due to their size, e.g., in terms of nodes and edges in the network.                                  &        \CheckedBox      &      \CheckedBox               \\
\bottomrule
\end{tabular}
\end{table}

\smallskip\noindent\textbf{Defining Dependencies}. 
The second set of perils (Perils 4-5) is related to dependency relationships between software units, and only the first case study is applicable. To address these perils, researchers should adopt the following strategy:

\begin{enumerate}
    \item Researchers should not rely solely on listed dependencies in configuration files (e.g., pom.xml, package.json, etc.) as a measure of dependency between two components. Instead, code-centric approaches should be used to validate which libraries are actually depended upon.
\end{enumerate}

For example, in the first case, in addition to mining the configuration information, the authors also analysed the similarity of the source code contributions to address Peril 4. Regarding Peril 5, since the study's objective was to investigate changes to the configuration files, the risk of the update not being executed was deemed less important.
It is important to note that the second case study did not include dependency analysis and, therefore, these perils did not apply.

\begin{figure*}[]
    \centering
    \begin{subfigure}{0.9\linewidth}
         \includegraphics[width=1.1\textwidth]{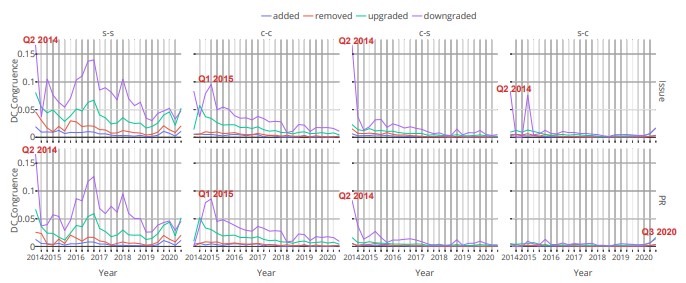}
         \caption{Visualization of a Time analysis for 107,242 libraries.}
     \end{subfigure}
     \begin{subfigure}{0.9\linewidth}
         \includegraphics[width=1\textwidth]{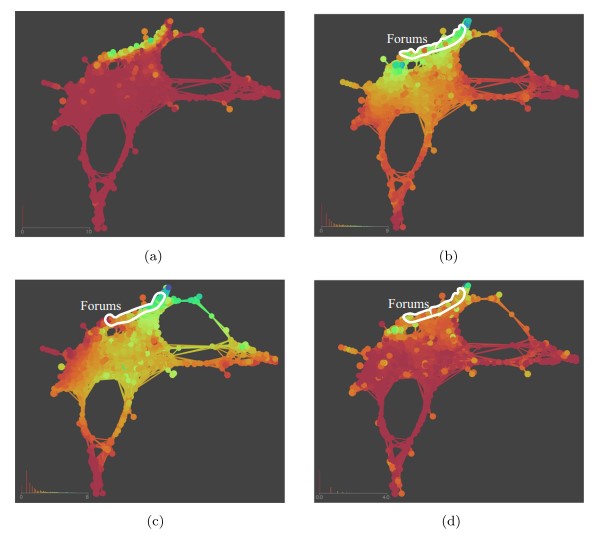}
         \caption{A Visual Topology map for 832,058 threads}
     \end{subfigure}
    \caption{Visualisation examples for the two case studies}
    \label{fig:visCase}
\end{figure*}

\smallskip\noindent\textbf{Defining Boundaries}.
The third set of perils (Perils 6-9) is related to the definition of boundaries and completeness and is relevant for both case studies. To mitigate these perils, we recommend the following strategies:

\begin{enumerate}
    \item Researchers should recognise that a dormant project does not necessarily mean that it is inactive. Instead, studies can use alternative heuristics, such as the number of dependents and dependencies, as better indicators of a project's importance in the ecosystem.
    \item Researchers should not rely solely on the programming language to define sub-communities. Using a common package manager for the programming language is a more effective rule of thumb for distinguishing boundaries.
    \item Researchers should avoid random sampling. Instead, sampling should be tailored to the research goals by considering factors such as an appropriate time window or focussing on specific attributes of components (e.g., most dependents, most popular, most contributors).
\end{enumerate}

Peril 6 did not apply to any of the case studies. 
Particularly for the first case, since the goal was to explore the npm package ecosystem, we assumed that the boundaries were clearly defined by the npm registry. 
Similarly, the second case study used the generic Eclipse platform as the boundary. 
Peril 7 was applied to the npm study, while Peril 8 was applied to both case studies. 
As a result, the two cases conducted a qualitative analysis of the dataset to gain deeper insights.
In the first case study, a three-month time window was created to capture dependencies. 
For the second case study, forum contributors were sampled into three groups (i.e., junior, member, or senior) according to the sliding window of their contributions. 

\smallskip\noindent\textbf{Visualisation}.
The final peril (Peril 10) relates to visualisation, which can be challenging due to the vast size and complexity of software ecosystems. As it is not feasible to visualise every aspect of an ecosystem simultaneously, a focused approach is necessary. A mitigation strategy is to select specific attributes of the ecosystem (e.g., the most dependent, most popular, and most contributions) that align with the research needs and objectives. 

Figure \ref{fig:visCase} shows two cases where visualizations are employed to gain insights, especially for large datasets.
In the first figure (a), we visualize the distributions of the data set and applied the appropriate statistical tests, along with the effect size, to test our hypotheses and answer research questions. 
In the second example (b), although not directly related to package ecosystems, the authors utilized a topological visualization \cite{Lum2013ExtractingIF} to gain insights on the over 800,000 forum threads of discussions.

\section{Chapter Summary}
In this chapter, we explore the various aspects of mining information from the software package ecosystem, presenting three promises and ten perils that researchers should be aware of when undertaking such tasks. The chapter is structured around four key processes for mining: 1) Planning what Information to Mine, 2) Defining Components and their Dependencies, 3) Defining Boundaries and Completeness, and 4) Analysing and Visualising the Data. To help new and experienced researchers navigate these challenges, we introduced the SUG model, which can serve as a valuable tool to minimise threats to validity. Although some perils may be more relevant to specific research objectives, our aim is to equip researchers with the knowledge and resources needed to confidently gather and integrate software package ecosystem data into their work.

\bibliographystyle{spmpsci}
\bibliography{book/references.bib}
\end{document}